# Sub-Mbps key-rate continuous-variable quantum key distribution with local-local-oscillator over 100 km fiber

Yaodi Pi[1], Heng Wang[1], Yan Pan[1], Yun Shao[1], Yang Li[1], Jie Yang[1,2], Yichen Zhang[2], Wei Huang[1†], and Bingjie Xu[1,2*]

[1]*Science and Technology on Communication Security Laboratory, Institute of Southwestern Communication, Chengdu 610041, China*
*\*Corresponding author: huangwei096505@aliyun.com and xbjpku@163.com*
[2] *State Key Laboratory of Information Photonics and Optical Communications, Beijing University of Posts and Telecommunications, Beijing 100876, China;*

We experimentally demonstrated a sub-Mbps key rate Gaussian-modulated coherent-state continuous-variable quantum key distribution (CV-QKD) over 100 km transmission distance. To efficiently control the excess noise, the quantum signal and the pilot tone are co-transmitted in fiber channel based on wide-band frequency and polarization multiplexing methods. Furthermore, a high-accuracy data-assisted time domain equalization algorithm is carefully designed to compensate the phase noise and polarization variation in low signal-to-noise ratio. The asymptotic secure key rate (SKR) of the demonstrated CV-QKD is experimentally evaluated to be 10.36 Mbps, 2.59 Mbps, and 0.69 Mbps over transmission distance of 50 km, 75 km, and 100 km, respectively. The experimental demonstrated CV-QKD system significantly improves transmission distance and SKR compared to the state-of-art GMCS CV-QKD experimental results, and shows the potential for long-distance and high-speed secure quantum key distribution. © 2022 Optical Society of America

Quantum key distribution (QKD) can provide an unconditional secure key for legitimate communication parties [1-3] over insecure channel, which can be mainly categorized into discrete-variable and continuous-variable schemes [4-8]. The continuous-variable quantum key distribution (CV-QKD) is considered to be a promising candidate for quantum secure network due to its high compatibility with classical optical communication and potential high secure key rate (SKR) within metropolitan distance [9,10]. Particularly, the Gaussian-modulated coherent-state (GMCS) CV-QKD protocol, as the most widely developed CV-QKD protocol, has made significant progress in both theoretical security proofs and experimental techniques in recent years [11-16].

Achieving high SKR and long transmission distance is of great importance for the practical application of CV-QKD, where an amount of developments have been reported recently based on GMCS protocols. Based on the transmitted local oscillator (TLO) scheme, the performance of 1 Mbps@25km [17] and 6.214 bps@202km [18] have been realized. However, the TLO scheme suffers from the crosstalk noise introduced by the strong TLO signal, and leaves some security loopholes which can be exploited by the eavesdropper to mount quantum hacking attacks [19-21]. Recently, the local local oscillator (LLO) scheme is proposed and successfully demonstrated, where the SKR of 26.9 Mbps@15km [22] and 7.04 Mbps@25km [23] have been achieved. Moreover, the transmission distance for GMCS CV-QKD with LLO scheme has been improved to 60km over fiber channel [24]. Up to now, it is still difficult to achieve a long-distance GMCS CV-QKD with LLO scheme. The restriction mainly comes from the crosstalk due to nonlinear effects, polarization variation, $X$ and $P$ quadrature imbalance and imperfect phase noise compensation. With the increase of system repetition rate, the above restriction will become more prominent.

In this paper, we present a sub-Mbps SKR experimental demonstration of LLO GMCS CV-QKD system over all transmission distance within 100 km. To improve the repetition rate of the system, an electro-optic IQ modulator is used to prepare the high-speed gaussian modulated coherent states. To efficiently control the excess noise, a pilot tone that co-transmitted with quantum signal is used to compensate for frequency and phase drifts, where wide-band frequency and polarization multiplexing method is adopted to decrease the crosstalk noise between them. Furthermore, a high-accuracy data-assisted time domain equalization algorithm is carefully designed to realize polarization variation correction, $X$ and $P$ quadrature imbalance compensation, and phase noise compensation in long transmission distance. Finally, a high performance LLO GMCS CV-QKD setup with high SKR and long distance is experimentally demonstrated, where the asymptotic SKR achieve to be 10.36 Mbps, 2.59 Mbps, and 0.69 Mbps over the transmission distance of 50 km, 75 km, and 100 km in standard fiber channel, respectively.

The diagram of the experimentally demonstrated LLO GMCS CV-QKD system is shown in Fig. 1. At Alice's side, the continuous optical wave with wavelength 1550.22 nm is divided into two branches by PBS1, which is originated from a laser (Laser1, NKT Photonic Basik) with narrow linewidth (<1 kHz) and low relative intensity noise (RIN < -100 dBc/Hz). In the upper branch, the continuous optical wave is sent into a commercial IQ modulator (Fujitsu FTM7962EP, 3-dB bandwidth: 22 GHz) to generate Gaussian modulated coherent states with 1Ghz repetition rate, where the electrical modulation signals are generated from an arbitrary waveform generator (Keysight AWGs 8195A) working at 30 GSa/s. Then, a variable optical attenuator (VOA1) is used to adjust the modulation variance of the output quantum signals. In the lower branch, VOA2 is used to adjust the intensity of the pilot tone adaptively. Then, the quantum signal and the pilot tone are combined by a polarization beam coupler (PBC) and co-transmitted over the fiber channel (SMF-28) at different distances of 50 km, 75 km, and 100 km, respectively.

At Bob's side, the quantum signal and the pilot tone are divided into two branches by PBS2. The frequency of the LO signal originated from the Laser2 (NKT Photonic Basik) is set to be 2 GHz offset from Laser1. The LO is divided into two branches by PBS3 for supplying sufficient optical power to realize shot-noise-limited coherent detection. Then, the quantum signal and pilot tone are coupled with their LO by the optical couplers (OCs) and detected by the balanced homodyne detectors (THORLABS PDB480C with 3-dB bandwidth 1.6 GHz), respectively. In addition, the detectors' electrical outputs are monitored and collected by a real-time oscilloscope working at 10 GSa/s (Keysight DSAZ254A) for the subsequent digital signal processing (DSP). In the DSP, the raw data of the quantum signal is well recovered to achieve a low level of excess noise for distilling the final SKR.

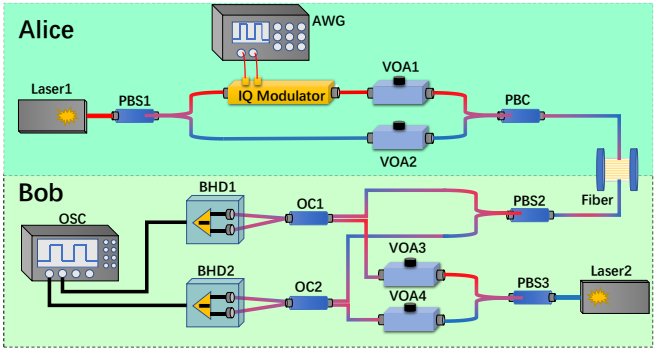

Fig. 1 Experimental setup of the proposed LLO GMCS CV-QKD system. AWG: arbitrary waveform generator, PBS: polarization beam splitter, PBC: polarization beam coupler, VOA: variable optical attenuator, OC: optical coupler, BHD: balanced homodyne detector, OSC: Oscilloscope.

The crosstalk between the pilot tone and the quantum signal is one of the main sources of excess noise, which needs to be eliminated by multiplexing methods such as frequency division multiplexing. Traditional frequency division multiplexing scheme usually simply separates the spectrum range of quantum signal and the pilot tone. However, the pilot tone may cause spontaneous Rayleigh or Brillouin scattering spectra in fiber channel, which will impose leaked photons to the quantum signal. In our CV-QKD setup, a wide-band frequency division is used to avoid the influence of the scattering spectra. As shown in Fig. 2, the frequency difference between the two lasers is set to be 2 GHz. Meanwhile, the center frequency of the detected quantum signal is set to be 700 MHz through a 1.3 GHz frequency shift generated by AWG. The setting of the frequency range can effectively reduce the crosstalk noise. Note that the bandwidth of the scattered noise is related to the intensity of the pilot tone, which should be reasonably set in real experiment.

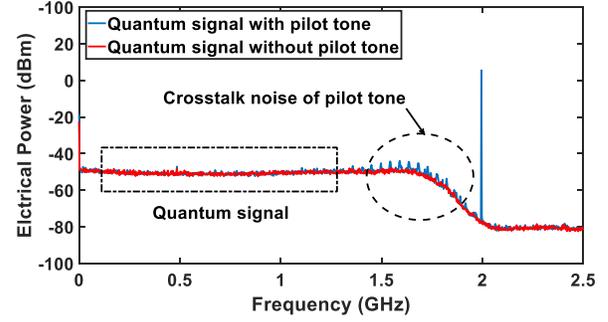

Fig. 2 Measured spectra of the quantum signal with crosstalk noise of pilot tone. The dotted box is the frequency domain range of the quantum signal. The dotted circle is the crosstalk noise of the pilot tone. The blue signal at 2 GHz is the beat signal of the pilot tone and local oscillator.

To further improve the performance of the long-distance CV-QKD system, phase noise, frequency offset, and polarization correction are fully digitally compensated by a carefully designed DSP algorithms. Here, the main DSP algorithms adopted in our experiment are shown in Fig. 3.

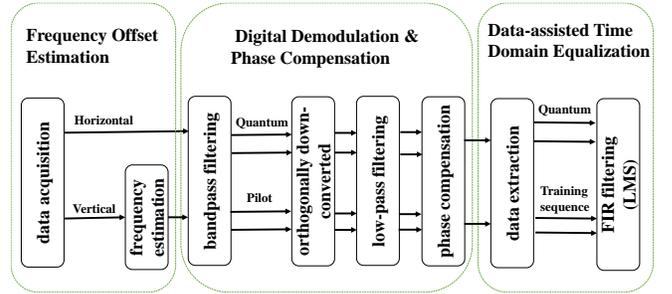

Fig. 3 The main DSP algorithms: 1) Frequency offset estimation; 2) Digital demodulation and phase compensation; 3) Data-assisted time domain equalization algorithm.

1) Frequency offset estimation. Due to the frequency drift of two lasers, the detected intermediate frequency (IF) is not stable which will affect the precise demodulation of the quantum signal. Therefore, frequency offset estimation is performed by the pilot tone in the frequency domain, and the estimated IF is used for digital X/P quadrature demodulation.

2) Digital demodulation and phase compensation. The quantum signal and pilot tone are filtered according to their desired bandpass filtering bandwidth, respectively. As mentioned previously, the desired bandpass filtering bandwidth of the quantum signal and pilot tone is set to be 1.3 GHz and 10 MHz. After the bandpass filtering, the quantum signal and the pilot tone are orthogonally down-converted and low-pass filtering to demodulate the X and P quadratures. The dominant phase noise can be compensated by sharing the phase of the pilot tone.

3) Data-assisted time domain equalization algorithm. With the help of the training sequence and least-mean-square (LMS) algorithm, a real-valued finite-impulse response (FIR) filter is implemented [25]. Assuming $s_{hx}$, $s_{hp}$, $s_{vx}$, and $s_{vp}$ are the X and P quadratures of horizontal and vertical polarization for the training sequence, respectively. The outputs of the real-valued finite-impulse response (FIR) filters are given as:

$$s^o_{hx} = \omega_{11}s_{hx} + \omega_{12}s_{hp} + \omega_{13}s_{vx} + \omega_{14}s_{vp}$$
$$s^o_{hp} = \omega_{21}s_{hx} + \omega_{22}s_{hp} + \omega_{23}s_{vx} + \omega_{24}s_{vp}, \quad (1)$$

where $\omega_{11}, \omega_{12}, \omega_{13}, \omega_{14}, \omega_{21}, \omega_{22}, \omega_{23}$, and $\omega_{24}$ are the tap weights of the FIR filters. Here, the LMS algorithm with pilot-aided is used to update the tap weights. After convergence of the training process, functions of the real-valued FIR filters can be achieved, which realizes polarization variation correction, X and P quadrature imbalance compensation, and residual phase noise compensation by performing the process of Eq. (1) to quantum signals. The distributions of X and P quadratures before and after DSP are shown in Fig. 4, and it can be seen that the correlation between Alice and Bob is improved after DSP.

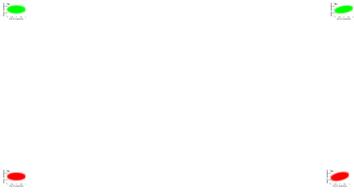

Fig. 4 Measured X quadrature and P quadrature (70 000 points) between Alice and Bob before and after DSP, (a) X quadrature before DSP;(b) X quadrature after DSP;(c) P quadrature before phase compensation (d) P quadrature after DSP.

The excess noise is measured on the block of size $1 \times 10^6$ with transmission distance of 50 km, 75 km, and 100 km, respectively, as shown in Fig. 5. The blue solid lines in Fig. 5 show the null key rate threshold of excess noise at different distances. The dashed line in the figure shows the average excess noises of the CV-QKD system in about two hours of continuous operation. The repetition rate of the modulated quantum signals is 1 GHz, the detection efficiency is 0.56, the electrical noise is 0.16 SNU, the modulation variance is 3.9 SNU, the reconciliation efficiency is 0.95, and the experimentally measured average of excess noise is around 0.039, 0.040 and 0.040 at transmission distance of 50 km, 75 km, and 100 km, respectively.

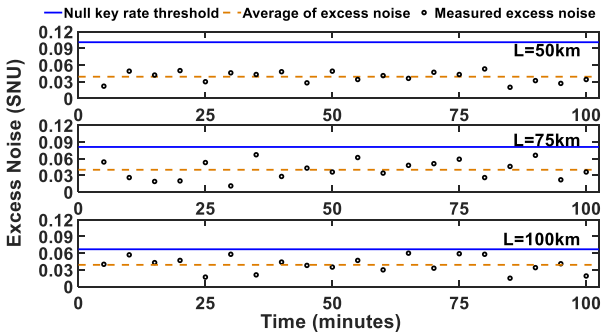

Fig. 5 Excess noise is measured in SNU within 100 minutes. The blue solid line represents the null key rate threshold of excess noise and the yellow dashed line represents the average excess noise.

Based on the measured excess noise, the asymptotic SKR of the setup can be evaluated as [26]:

$$SKR = R_s(\beta I_{AB} - \chi_{EB}), \quad (2)$$

where $R_s$ is the repetition rate of the system, $\beta$ is the reconciliation efficiency [27, 28], $I_{AB}$ is the mutual information between Alice and Bob, and $\chi_{EB}$ is the Holevo bound between Eve and Bob. The corresponding SKR with experimentally measured excess noise in Fig. 5 is shown in Fig. 6, where the calculated average key rate is represented with a dashed line.

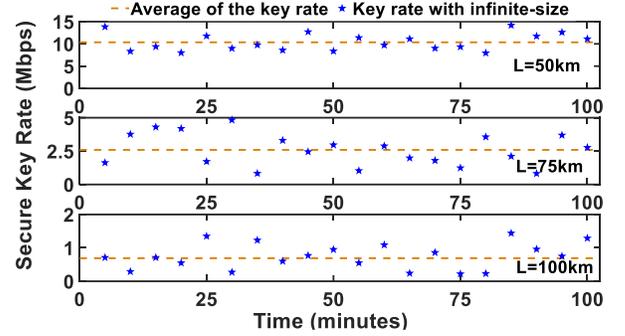

Fig. 6 SKR corresponding to each excess noise in Fig. 5. The yellow dashed line represents the average key rate.

In addition, the previously reported experimental results are also shown in Fig. 7 for comparison. The red solid line represents asymptotic SKRs at different distances in our experiment. The remaining solid lines and symbols respectively represent the experimental SKRs obtained from the corresponding references [18, 23, 24, 29, 30, 31]. The average asymptotic SKR in our experiment achieves 10.36 Mbps, 2.59 Mbps, and 0.69 Mbps over the transmission distance of 50 km, 75 km, and 100 km, respectively. Compared with the state-of-art results in this field, our experimental results improve the SKR and transmission distance of CV-QKD.

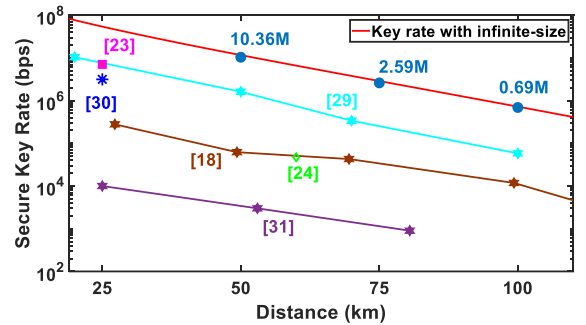

Fig. 7 SKR versus secure transmission distance curves. The red solid line represents infinite-size SKRs at different distances in our experiment. The remaining solid lines and symbols respectively represent the experimental SKRs obtained from the corresponding references [18, 23, 24, 29, 30, 31].

In this paper, we experimentally demonstrated a sub-Mbps key rate GMCS CV-QKD over 100 km transmission distance. The achieved SKRs are demonstrated at different distances, including 50 km, 75 km, and 100 km, up to be 10.36 Mbps, 2.59 Mbps, and 0.69 Mbps, which are higher than state-of-the-art GMCS CV-QKD experimental results. Notably, to achieve a long time stable SKR performance for practical system, real-time accurate shot noise estimation, precise frequency, and polarization variation compensation are required. Furthermore, limited by the effective block size of the setup, the SKR will be decreased significantly if the finite-size effect or composable security is considered. However, with the deepening of investigation (i.e. real-time shot noise calibration, novel DSP algorithms, larger block size, and advanced high performance post-processing), we believe the mentioned challenges will be overcame, and the proposed LLO CV-QKD scheme can be effectively applied in practical secure scenarios in long-distance in the future.

**Funding.** The National Science Foundation of China (Grants No. 62101516, No. 62171418, No. 61771439, No. U19A2076, and No. 61901425), the Chengdu Major Science and Technology Innovation Program (2021-YF08-00040-GX), the Technology Innovation and Development Foundation of China Cyber Security (Grants No. JSCX2021JC001), the National Key Research and Development Program of China (2020YFA030970X), the Sichuan Application and Basic Research Funds (Grants No. 2021YJ0313), the Chengdu Key Research and Development Support Program (2021-YF05-02430-GX, 2021-YF09-00116-GX), the Sichuan Science and Technology Program (Grants No. 2019JDJ0060, No. 2020YFG0289, and No. 2022YFG0330) and the National Science Key Lab Fund Project (6142103200105), the Major Project of the Department of Science and Technology of Sichuan(2022ZDZX0009), foundation of Science and Technology on Communication Security Laboratory (Grant No. 61421030402012111).

**Disclosures.** The authors declare no conflicts of interest.

**Data availability.** Data underlying the results presented in this paper are not publicly available at this time but may be obtained from the authors upon reasonable request.